\UseRawInputEncoding
\documentclass[10pt,letterpaper]{article}

\usepackage[top=0.85in,left=2.75in,footskip=0.75in,marginparwidth=2in]{geometry}

\usepackage[utf8]{inputenc}

\usepackage{cite}

\usepackage{nameref,hyperref}

\usepackage{booktabs}

\usepackage{microtype}
\DisableLigatures[f]{encoding = *, family = * }

\raggedright
\usepackage{changepage}

\usepackage[aboveskip=1pt,labelfont=bf,labelsep=period,singlelinecheck=off]{caption}

\makeatletter
\renewcommand{\@biblabel}[1]{\quad#1.}
\makeatother

\usepackage{lastpage,fancyhdr,graphicx}
\usepackage{epstopdf}
\fancyheadoffset[L]{2.25in}
\fancyfootoffset[L]{2.25in}

\usepackage{color}

\definecolor{Gray}{gray}{.25}

\usepackage{graphicx}

\usepackage{sidecap}

\usepackage{wrapfig}
\usepackage[pscoord]{eso-pic}
\usepackage[fulladjust]{marginnote}
\reversemarginpar

\begin{document}
\vspace*{0.35in}

% title goes here:
\begin{flushleft}
{\Large
\textbf\newline{Teacher, But Also Student: Challenges and Tech Needs of Adult Braille Learners with Sight}
}
\newline
% authors go here:
\\
Quan Zhou, Cameron Cassidy, Alyson Yin, Stacy Branham
\\
\bigskip
\bf University of California, Irvine
\\
\bigskip
* quanz13@uci.edu, camcass@uci.edu, yuehany1@uci.edu, sbranham@uci.edu

\end{flushleft}

\section*{Abstract}
Braille literacy is critical for blind individuals’ independence and quality of life, yet literacy rates continue to decline. Though braille instructors in integrated K-12 classrooms play a central role in literacy development in blind youth, prior research on braille learning almost exclusively focuses on \textit{blind adolescent students}. As a result, we still know little about how \textit{sighted adult teachers} learn braille. To address this, we interviewed 14 educators, including 13 certificated Teachers of Students with Visual Impairments (TVIs) and 1 paraeducator, who learned braille as adults. We found that they: (1) lack consistent braille exposure to reinforce knowledge and skill; (2) have limited time to practice due to myriad responsibilities of adulthood; and thus, (3) seek learning tools that are engaging and efficient. Our research draws attention to the needs of a group of braille learners who have been overlooked and identifies new design opportunities to facilitate braille literacy.

% now start line numbers
% \linenumbers

\section{Introduction}
Among the roughly 1 million people with blindness in the United States \cite{CDC2024}, braille literacy has experienced a dramatic decline over the past half century. In 1960, more than half of blind students in the U.S. were braille literate \cite{Brailleworks2019}, but today that number has dropped to around 10\% \cite{Schroeder2007, Gadiraju2020}. This decline has been attributed to the increasing availability of digital assistive technologies such as screen readers \cite{WUFT2016}, as well as systemic challenges such as \textit{"mainstreaming"} blind children into general education classrooms \cite{NBPNeedforBraille}. The potential consequences of these shifts, however, are significant: one study found that congenitally blind adults who were taught braille as their original reading medium were employed at more than twice the rate of those who had learned to read using print-based alternatives, such as large print \cite{Ryles1996}. Moreover, roughly half of blind students drop out before graduating high school, in part due to extremely low literacy rates among this population \cite{Brailleworks2019}. 

Sustaining and expanding braille literacy requires teachers who support blind students. Teachers of Students with Visual Impairments (TVIs) and paraeducators fill this role in primary and secondary schools around the United States. TVIs are responsible for, among teaching many other accessibility-related skills, ensuring the students' proficiency in braille code. Beyond direct instruction, TVIs and paraeducators are also responsible for adapting classroom materials into braille and preparing other resources that make learning possible \cite{APH_TVI_role, Texas}. Yet, in HCI, the vast majority of research regarding braille learning focuses on younger learners, such as visually impaired children \cite{Lang2023, Sánchez2016, Milne2014, Ara2016, Nahar2015}, with a smaller set focusing on blind and low vision people across age ranges\cite{Forcelini2018, Wagh2016}.  
%However, many TVIs are sighted and must acquire braille proficiency as an adult, a process which, we argue, presents its own challenges that can limit the depth of instruction they are able to provide.

TVIs, as (primarily) sighted adult learners of braille, have largely been excluded from prior work. The few studies that have considered TVIs’ braille learning experiences are outside the HCI community. Herzberg et al.'s work explored previously braille-literate TVIs’ experience in switching from one braille code to another, documenting their approaches in learning new braille codes \cite{Herzberg2023}. In a study of a 9-month braille training program for TVIs and other sighted adults, Bola et al. \cite{Bola2016} reported sighted adults’ ability to learn braille tactilely. However, these studies focused on tactile learning rather than learning with sight; excluded learning contexts outside of formal training programs; and lacked contributions on TVIs’ unique technological needs for braille learning. To fill this gap, we asked the following research questions:
\begin{itemize}
    \item \textbf{RQ1:} What challenges do TVIs encounter when learning braille?
    \item \textbf{RQ2:} What tools and techniques do TVIs use to learn braille?
    \item \textbf{RQ3:} How do TVIs imagine technology better supporting their braille learning process?
\end{itemize}

To address these questions, we conducted semi-structured interviews with 14 educators who learn braille as adults by sight (13 TVIs and 1 paraeducator), focusing on their braille learning practices, challenges, and tech needs. Our analysis yielded three principal themes. First, sighted adult braille learners had advantages from previous life experiences while facing unique challenges from adult life. Second, TVIs applied various approaches to learn braille but still found learning resources inadequate. Third, TVIs envisioned specialized technologies and features designed for braille learning to utilize fragmented time and keep them engaged in braille learning long after formal education has been completed. We contribute (1) an empirical account of the challenges and strategies of sighted adult braille learners, contrasting their experiences with adolescents who learn braille; (2) a theoretical connection between these challenges and micro- \cite{Gassler2004} and wait-learning \cite{Carrie2017} frameworks from foreign language learning \cite{Carrie2017}, framing adult braille learning as a distributed, opportunistic practice; and (3) design directions for braille learning technologies that leverage everyday contexts, fragmented time, and para-social support to sustain motivation and retention.

\section{Background}
\subsection{Braille}
\begin{figure}[hbt!]
    \centering
    \includegraphics[width=\columnwidth]{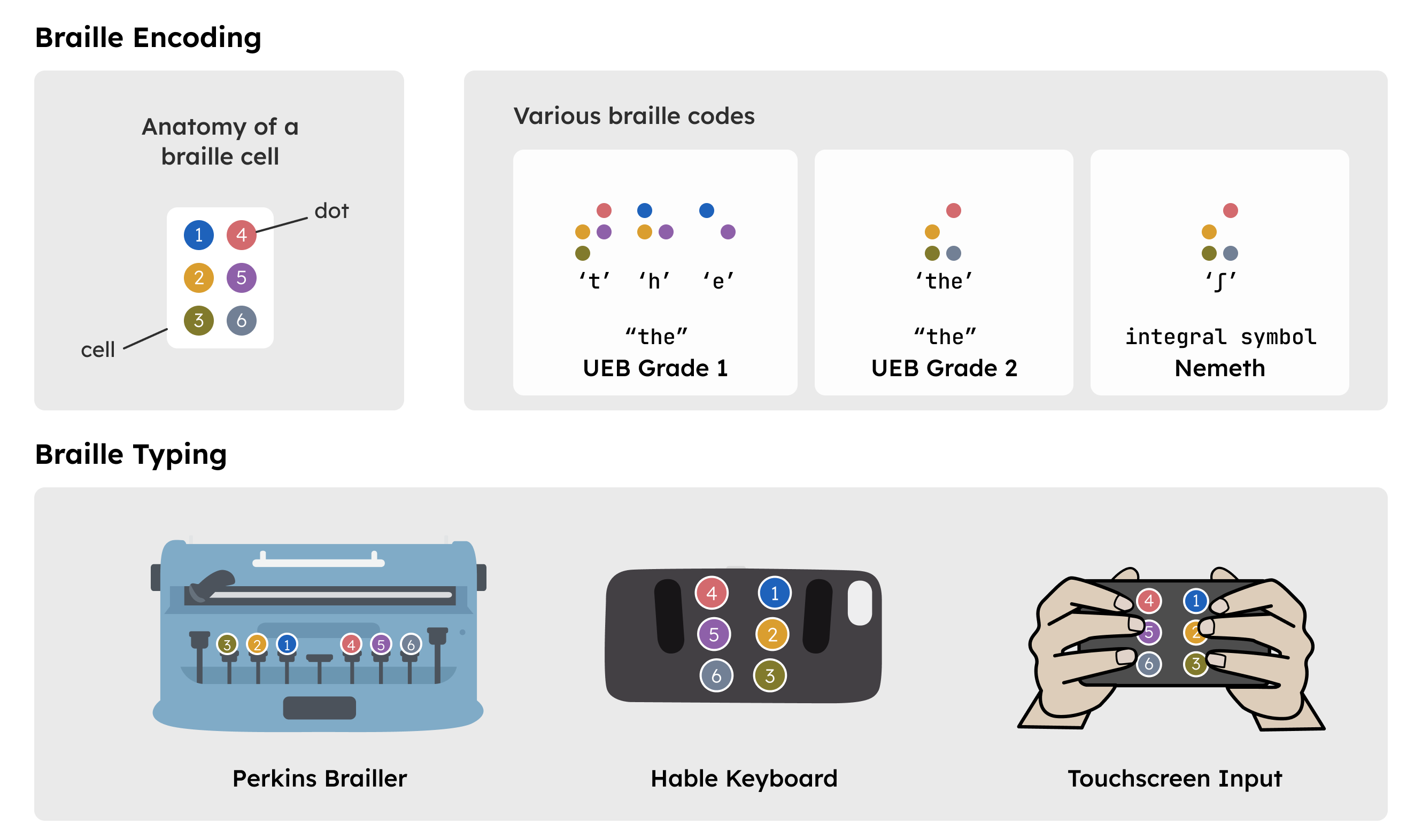}
    \caption{Anatomy of a braille cell, braille encoding, and typing methods. The top panel shows the structure of a braille cell and how the same dots are used differently in UEB Grade 1 (“the” spelled out), UEB Grade 2 (“the” contracted), and Nemeth (integral symbol). The bottom panel illustrates three common braille input devices with dots mapping: the Perkins Brailler, the Hable Keyboard, and touchscreen braille input on a smartphone}
    \label{fig:BackgroundFigure}
\end{figure}

Braille is a tactile reading and writing system for blind people invented in 1824 by Luis Braille \cite{Britannica2025}. The basic unit of the braille system is called a braille cell, formed by six raised dots, arranged in two columns and three rows (Fig. \ref{fig:BackgroundFigure}) \cite{Britannica2025}. While braille is not a language \cite{BrailleWorks2017}, encoding varies across countries, languages, usage context (e.g., mathematics, music, literary), and time \cite{Britannica2025}. Unified English Braille (UEB) is the mainstream braille system for literary braille in the United States \cite{ICEBXXXX, BrailleBug}. However, certain educational subjects might apply alternative systems. For example, non-English languages \cite{BrailleWorks2017}, mathematics and physical sciences (Nemeth and UEB \cite{ICEBXXXX, Bana2023, BrailleBug}), and music \cite{Bana2015} are specified in distinct braille codes. Literary UEB can be further broken down into non-contracted grade-1 braille, which interprets each letter by one braille cell, and contracted grade-2 braille, which applies contractions to reduce the length of code (Fig. \ref{fig:BackgroundFigure}), saving space and time depending on the users' braille fluency \cite{Britannica2025, ICEBXXXX}.

Tools for creating and accessing braille encompass manual, digital, and print-based approaches. A simple but widely used manual, portable tool is the slate and stylus \cite{slate_and_stylus_wiki}, which lets users punch braille dots one-by-one onto paper, in aligned rows. The Perkins Brailler, a braille “typewriter” \cite{Perkins_Brailler_wiki}, allows more efficient braille typing using a mechanical 6-key chorded keyboard in a large, heavy, metal chassis (Fig. \ref{fig:BackgroundFigure}). Digital technologies include refreshable braille displays, braille keyboards, and braille embossers \cite{NECO2025}. These technologies are not limited to a single form. For example, Hable markets a simple wireless braille keyboard for use with smartphones and tablets that accepts input via six physical buttons (Fig. \ref{fig:BackgroundFigure}) \cite{Hable}. 
%replace braille learning part by website content rather than researches
Tactile braille learners start by building pre-braille skills, which develops one's fingertip sensitivity, in order to perceive the dots comprising distinct braille cells \cite{Jurnal2021, Steinman2006}. However, those learning with full or partial vision may choose to skip this process and learn braille by sight. Next, learners progress through memorizing single letters, symbols, contractions, and finally move on to reading complete sentences and paragraphs \cite{Steinman2006}.

% Learning strategies that can be used include word instruction vocabulary, oral braille reading decoding, stimulus equivalence and constant time delay, while flashcard is a frequently used media \cite{Susanti2019}. 

%There are multiple resources for people to access braille learning materials. According to the Texas School for the Blind and Visually Impaired, braille resources include "tutorials for teachers; information on braille code and transcription; tactile graphic production and braille products; and information on braille instruction and curriculum" \cite{Tbr}. There are instructional devices with six-key braille input, audio feedback, and game features to engage young braille learners in learning and practicing, such as BrailleBuzz \cite{braillebuzz} and Polly \cite{polly}.

\subsection{Teacher of Students with Visual Impairments (TVI)}
A TVI is a certified educator with specialized training to provide instruction that meets visually impaired students’ unique educational needs. They serve students ranging in age from birth through 21 years old, depending on specific certification requirements, which vary by state in the USA \cite{tsbvi_TVI,tsvi_TVI}. TVIs are usually "itinerant", traveling between schools in a district to work with assigned students at each school. They may also meet with students before or after regular school classes \cite{APH_TVI_role}. TVIs have a variety of roles and responsibilities. Working with visually impaired students, TVIs' roles primarily include teaching the specific skills that they need due to visual impairment, preparing or obtaining accessible learning materials, conducting and assisting assessments. TVIs are also responsible for coordinating with others, such as students' family members, regular education teachers, paraeducators \footnote{Paraeducators (also known as paraprofessionals, teacher aides, or instructional aides) are primarily employed to assist students and their certificated educators in special education. While paraeducators' definitions and requirements varies across states, here are California's requirements towards paraeducators: https://www.cde.ca.gov/ci/pl/paraprofessionals.asp}, and other school personnel to discuss learning progress, suggest approaches on accessibility accommodations, and provide consultation and training for guiding the students \cite{APH_TVI_role, Texas}. 

\section{Related Work}
\subsection{Learning Technologies}
\subsubsection{Braille Learning}
There has been a long-standing interest in accessible technology design for braille learning. Considering the needs of people who are totally blind or have some usable vision, a variety of interactive modalities have been explored and evaluated, with a particular focus on tactility. Tangible User Interfaces, for example, have been widely explored for their potential in enhancing tactile braille learning. These designs often take the form of multiple blocks with enlarged braille characters placed next to each other for users, especially children, to learn basic letters and words in braille \cite{Lang2023, Sánchez2016}.  

Braille learning technologies tend to adopt various modalities for interaction. Tactile elements are often paired with audio output, as an additional modality to elicit attention and provide feedback \cite{Baqai2021, Forcelini2018, Wagh2016}, and audio input, for accessible voice control \cite{Wagh2016}. For example, BrailleBlocks combined tangible blocks and audio feedback with game features to support collaborative learning between parents and visually impaired children \cite{Gadiraju2020}.
%E-Braille introduced a braille kit with a braille keypad and audio input and output for early stage braille self-learning \cite{Wagh2016}.

Several braille learning apps explore mobile platforms and integrate gamification, to increase access and engagement. Mobile phone is frequently considered as a platform for learning and practicing braille, due to its portability and additional sensors (e.g., touchscreen, vibration motor) to support multi-modal interaction \cite{Milne2014}. Several mobile phone games have been designed for braille learning, such as BraillePlay, a smartphone game for blind children to learn braille characters \cite{Milne2014}; GBraille, an audio-based mobile phone game to support braille learning in English, Portuguese, and Spanish \cite{Ara2016}; and MBraille, a Bangla and English braille learning platform \cite{Nahar2015}. A study of BrailleBlocks \cite{Gadiraju2020} found that gamification proved not only engaging for children, but also motivated sighted adults, primarily the parents, to practice braille. 

However, most of these technologies were designed for visually impaired children or adolescents \cite{Lang2023, Sánchez2016, Milne2014, Ara2016, Nahar2015}, while very few have been designed for PWVIs across age ranges \cite{Forcelini2018, Wagh2016}. To the best of our knowledge, sighted adults have only been included in academic studies of braille learning technologies in their limited capacity as facilitators of children's braille learning \cite{Gadiraju2020}, rather than braille learners who also seek learning technologies. 

\subsubsection{Mobile-Assisted Language Learning} 
Even though braille is a code, distinct from a language, there are several qualities of braille learning that are akin to language learning—coding and decoding, learning contractions, formatting conventions, etc. We therefore incorporate a synthesis of related work regarding language learning technologies. 

There is growing interest in educational technology design for technology-assisted language learning. There are AI-assisted \cite{Liang2025, Li2024}, robot-assisted \cite{Nomoto2022, Randall2019}, computer-assisted \cite{Tu2024, Kremenska2007}, and mobile-assisted \cite{Hu2021, Zheng2024, Hu2023, Sandberg2011} learning tools and approaches. Among them, mobile-assisted technologies have been claimed by Hanif \cite{Hanif2020} as an effective way to facilitate language learning in combination with self-regulated learning—which involves self-motivation, self-evaluation, and self-regulation of the knowledge acquiring process \cite{ZimmermanandSchunk2011}. Several studies have been conducted towards mobile-assisted learning technologies, either evaluating existing tools\cite{Hu2021, Zheng2024, Hu2023, Muckenhumer2023} or new designs \cite{Sandberg2011}, confirming their effectiveness in assisting language learning, motivating students to practice using spare time \cite{Sandberg2011}, and enhancing individual learning demands such as adjusting learning pace \cite{Hu2021}. Combining technology-assisted learning with accessibility, Rahman et al. proposed a web-based platform to help children's sign language learning, in which participants expressed a preference for both in-class, collaborative activities and individual learning with the tool \cite{Rahman2025}.

% mobile assisted language learning is an effective augmentation of formal school learning, which motivating students to learn and practice using their spare time. \cite{Sandberg2011}
% utilized instructional online learning tools to enhance English listening, comfirm the resilience of mobile-assisted technology in learning and enhance individual learning demands such as adjusting learning pace.\cite{Hu2021}

% Pintrich, Zimmerman and Schunk suggested that SRL involves self-motivation, self-evaluation, and self-regulation of learners’ whole knowledge acquiring process. Correspondingly, learners would supervise their learning behavior and evaluate their learning outcomes \cite{ZimmermanandSchunk2011}.
%Hu2023: Analyze Duolingo; daily practice helpful in mobile assisted language learning

\subsection{Research on TVIs}
\subsubsection{TVI's Role in Visually Impaired Students' Literacy Development}
TVIs play a vital role in visually impaired students' braille development. Specifically, they are responsible for ensuring the students' proficiency in braille code to access all curricula, including literary- and STEM-related content. They also need to translate learning materials into braille and guide paraeducators in preparing braille materials and supporting students' braille learning \cite{Herzberg_Rett_2023, APH_TVI_role, Texas}. 

However, TVIs are not simply tasked with teaching a code. Rather, they are teaching visually impaired children how to read and write, building the literacy foundation to "create strong, motivated readers." This points to the significance of TVI's involvement in every aspect of students' braille literacy development, from learning the basic braille code and reading skills, to "fluency, vocabulary development, and comprehension skills." Beginners require years of one-on-one instruction with TVIs to build solid braille skills and a positive mindset towards reading and writing, which will facilitate their full participation in mainstream literacy instruction \cite{Swenson2008}. 
%This leads to the need of TVIs' own proficiency in braille.
% [add citation to: 1) TVIs' braille proficiency is important; 2) TVIs' braille proficiency is low -> calls for professional training]

\subsubsection{Sighted Adult Braille Learners}
Considering TVIs often learn braille as adults, we look to research on braille learning with vision or during the process of aging with vision loss. Studies have been conducted on facilitators and barriers faced by older adults with acquired vision loss as they learn braille \cite{Martiniello2020}. D'Andrea identified motivations of various sighted adults learning braille, including parents reading with blind children; print readers writing braille letters to blind friends and family members; and rehabilitation teachers, TVIs, and transcribers working towards certifications and creating accessible paperwork for their students \cite{Andrea1996}. Bola et al. held a 9-month program to teach sighted adults braille, in which 14 of 29 participants were braille teachers. Their study revealed sighted adults’ ability to learn braille tactilely, although with a much slower reading speed than visually impaired children \cite{Bola2016}. Despite decades of research, we have been unable to identify substantive scholarship on TVIs' experience of learning braille with sight. 

\subsubsection{HCI Research on TVIs}
Within Human-Computer Interaction (HCI), TVIs and other vision-related instructors are primarily studied as regards their instructional role with visually impaired students. Regarding school curriculum, TVIs are included in research on: accessibility setting of STEM courses \cite{Stefik2019, Huff2021, Mountapmbeme2021, Varghese2021}, technology designs to enhance visually impaired students' learning experience \cite{Shi2019, Poddar2024, Gadiraju2021}, strategies to support certain groups of visually impaired students, such as those with cortical/cerebral visual impairment \cite{Smolansky2024}, and perspective and usage towards teaching materials and technologies \cite{phutane_tactile_2022, Boadi-Agyemang2023, Baker2019}. Regarding braille learning, Martillano et al.'s study involved TVIs in testing a portable, low-cost single braille cell display they designed to assist braille teaching \cite{Martillano2018}. While those studies acknowledged TVIs' significant role in educating visually impaired students in both school curriculum and braille, there is a lack of research on TVIs' own challenges and needs as sighted adult braille learners.
\section{Methods}
\subsection{Participants}
We recruited 14 educators who identify as adult braille learners with full or partial sight (Table \ref{tab:participants}). Participants were recruited through the existing social networks of the authors, followed by snowball sampling. All participants are native English speakers from the United States. Participants were between the ages of 26 and 74. The number of years they have been learning braille varies greatly, ranging from 1 to 55. Participants also have various teaching experiences, including being a paraeducator, braille transcriber, and TVI. It is worth noting that P5 is a braille teacher in several TVI credentialing programs across the United States, and P14 assists her grading students' braille assignments in one such program.

\subsection{Procedure}
We conducted semi-structured, audio-recorded interviews with 14 educators over Zoom. All interviews were conducted in English by the first author between June 2025 and August 2025. The interviews lasted between 45 and 65 minutes, with an average of 60 minutes. Prior to the interviews, we emailed participants a pre-interview survey, collecting their basic demographic information regarding braille learning experience. We acquired verbal consent at the beginning of the interview. During the interview, questions covered topics such as motivations for learning braille, challenges encountered, learning strategies and tools adopted, and potential features for novel braille learning technologies. Participants were compensated for their time at a rate of \$20 per hour in the form of a gift card. The study was approved by the authors’ institutional review board (IRB).

\subsection{Data Analysis}
All audio-recorded interviews were auto-transcribed and manually edited by researchers for quality and accuracy. We then conducted a thematic analysis of transcripts \cite{Braun_Clarke_2006b}. The first, second, and third authors independently open coded ten randomly selected transcripts; met weekly to discuss emerging themes and refine codes; and created a codebook with codes such as "learning braille by sight", "having limited practicing time", and "desiring gamification features." After that, the first author coded the remaining 4 transcripts individually, and discussed the generated codes with the team. The first author iteratively re-analyzed prior coded transcripts when new themes were constructed. We have arranged the headings and subheadings of the Findings section to correspond to the higher-level and lower-level themes we generated from coding. 

% \begingroup
% \renewcommand{\arraystretch}{1.2}  
%%% Uncomment 2 lines above and the line right below \end{table} to adjust line heights %%%
\begin{table}
    \begin{adjustwidth}{-1.5in}{0in}
    \caption{Participant demographics}
    \label{tab:participants}
    \begin{tabular}{l c c p{12em} p{9em} c} % value in parenthesis defines column width
        \toprule
        \textbf{ID} & \textbf{Age} & \textbf{Gender} & \textbf{Self-reported visual ability} & \textbf{Job / Experience} & \textbf{Years since learned braille}\\ \midrule
         P1 & 35& Man& Fully Sighted& Paraeducator, Braille Transcriber& 10\\ \midrule
         P2 & 29& Man& Fully Sighted& Paraeducator, TVI& 4\\ \midrule
         P3 & 34& Man& Fully Sighted& TVI& 8\\ \midrule
         P4 & 28& Woman& Fully Sighted& TVI& 3\\ \midrule
         P5 & 51& Woman& Fully Sighted& TVI, Braille Teacher for TVIs& 15\\ \midrule
         P6 & 48& Man& Fully Sighted& TVI& 1\\ \midrule
         P7 & 44& Woman& Fully Sighted& Paraeducator, Braille Transcriber& 4\\ \midrule
         P8 & 54& Woman& Fully Sighted& Paraeducator, TVI& 1\\ \midrule
         P9 & 56& Woman& Fully Sighted& Paraeducator, TVI& 7\\ \midrule
         P10 & 55& Woman& Fully Sighted, start to have macular degeneration& TVI& 1\\ \midrule
         P11 & 26& Woman& Low Vision& TVI& 1\\ \midrule
         P12 & 32& Woman& Fully Sighted& TVI& 2\\ \midrule
         P13 & 53& Man& Fully Sighted& TVI& 3\\ \midrule
         P14 & 74& Woman& Fully Sighted (while learning braille), identified as cataracts pre surgery before interview& TVI, Braille Assignments Grader for TVIs& 55\\
         \bottomrule
    \end{tabular}
    \end{adjustwidth}
\end{table}
% \endgroup

\section{Findings}

\subsection{Adult Braille Learners with Sight}
\begin{figure}[h]
    \centering
    \includegraphics[width=\columnwidth]{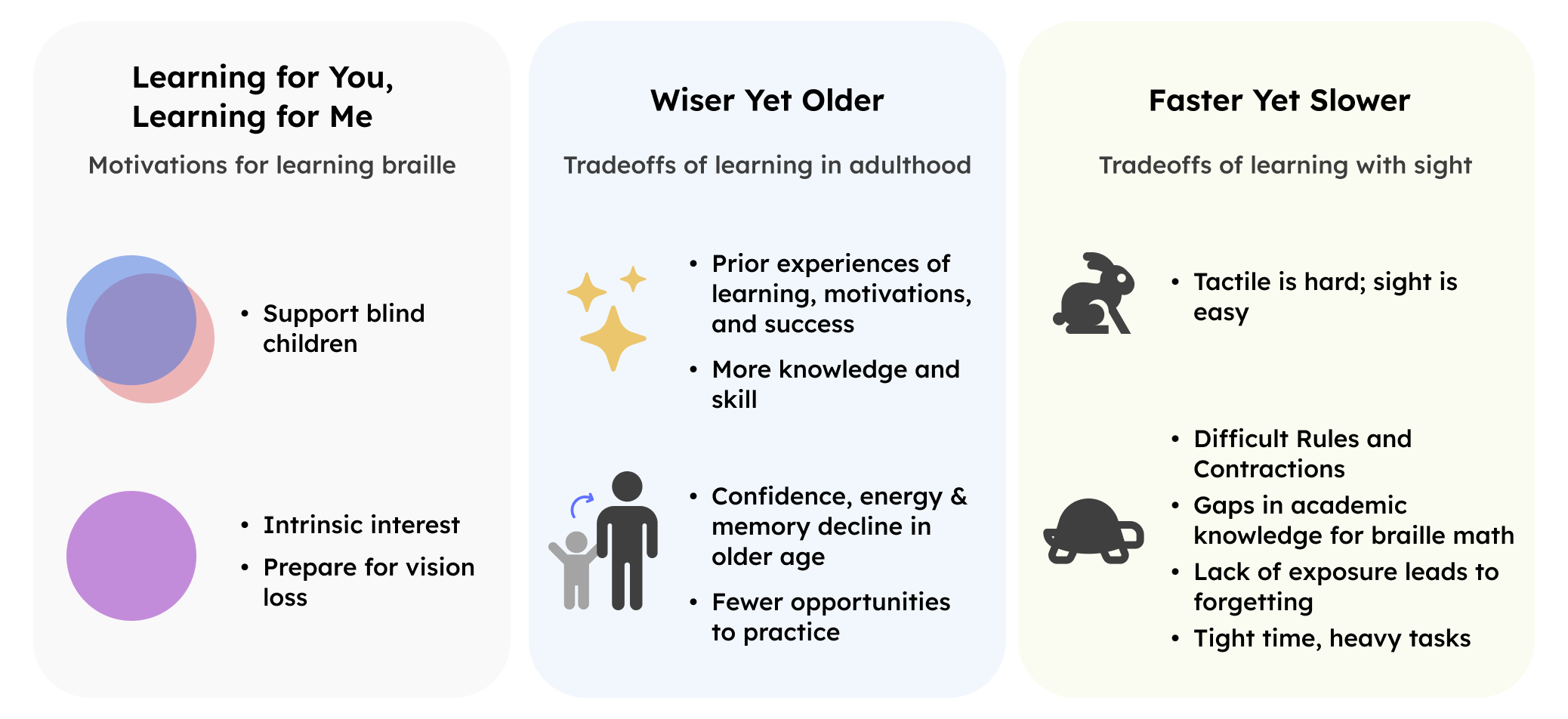}
    \caption{ Summary of three themes from Section 5.1: Learning for you, Learning for Me—motivations for learning braille, Wiser Yet Older—tradeoffs of learning in adulthood, and Faster Yet Slower—tradeoffs of learning with sight.}
    \label{fig:ThematicStructure}
\end{figure}

Sighted adult TVIs had a range of motivations to study braille, a learning experience that was shaped by the dual challenges of juggling adult responsibilities and aging, as well as the tradeoffs between visual and tactile learning (Fig. \ref{fig:ThematicStructure}).
\subsubsection{Learning for You, Learning for Me}
TVIs were motivated to learn braille by a desire to support their blind students and also themselves. All participants (n = 14) primarily learned braille for teaching purposes. Among those, nine participants identified the need of meeting the requirements of TVI training certifications, and the rest learned braille on-the-job as paraeducators to teach their blind students braille (P1, P2, P7, P9, P13). Other teaching-related motivations included: preparing for students with progressive vision loss (P6), preparing for future braille students (P3, P12, P13, P14), and being a role model for other older adults learning braille (P10).

Participants also learned braille for their personal development (n = 5). Four participants identified their intrinsic interest in braille (P8, P9, P12, P14). P10 learned braille to get prepared for her own vision loss due to macular degeneration: 
\begin{quote}
\textit{“I don’t want to feel limited. I would rather be set-up to be independent and fully functioning... to feel adequate instead of inadequate."} (P10) 
\end{quote}

\subsubsection{Learning Braille as an Adult: Wiser Yet Older}
TVIs reported many advantages and drawbacks of learning braille as an \textbf{adult}, such as prior knowledge and maturity. However, the realities of adult life and challenges of aging introduced tensions to their learning. 

%wiser
Adult learners benefited from years of formal education and mature learning strategies. Participants (n = 7) reported that, unlike children, they could draw on prior knowledge and skills, which they believed accelerated their braille learning. Prior experiences of reading and writing in print were considered especially helpful (n=9), as TVIs were then \textit{"just learning the code"} rather than \textit{"learning how to read"}, making braille learning \textit{"faster"} (P9). 

In addition, adult learners approached learning with greater maturity (P6, P9, P10). They were intrinsically motivated, \textit{“just doing it strictly to learn”} (P9), determined to \textit{"just get through it"} (P10), and \textit{"didn't have to be entertained"}. Adults also had confidence in their ability to achieve, as they \textit{"had successes in other spaces and places”} (P6). Adults also knew how to \textit{"advocate"} (P8) for themselves and \textit{"get that access"} to the tools and resources needed to succeed. Last, adults reported having \textit{"a growth mindset...a positive mindset, which helps a lot"} (P6) in their learning.

Adult learners, though able to set their own schedules, often struggled to prioritize braille learning and practice because of competing responsibilities (n=9). Many cited work (n=6) and childcare (n=2) as barriers, leaving them with limited time and energy:
\begin{quote}
\textit{“[Learning after work is] not something that you can really do. Your brain is already tired, and then you’re trying to learn all this new information. And having the time; there’s no such thing as enough time"} (P13).
\end{quote}
Although adults \textit{“have the liberty of making our own schedule”} (P8), braille often became the \textit{“last priority”} (P9) or just an \textit{“extra thing”} (P6). As a result, most adult learners (n=7) reported fewer opportunities to practice regularly.

%older
Participants also reported that the natural declines related to aging, both physical and mental, introduced challenges. For some (n=2), this was simply a decrease in energy:
\begin{quote}
    \textit{"...when you start reaching your fifties, I will be honest in that you get fatigued earlier in the evening. You can’t stay up late, and so you only have so much learning time after you’re back home from work, and I found that very limiting."} (P10)
\end{quote}
Adult learners also reported that \textit{"young children’s brains are much more pliable and able to absorb new words, contractions, and skills"} (P10), whereas \textit{“things get a little bit harder to remember and learn”} (P13) in adulthood. As a result, some participants held a negative self-perception of learning in older age, even at the same time as they held all the previously mentioned self-motivations and positive growth mindsets (P6, P10). P10 said \textit{“I’m too old for this"} twice, questioning \textit{"how am I gonna learn [braille]"}, while P6 thought, \textit{“It’s not quite ‘teaching old dogs new tricks’, but it kinda is.”}

\subsubsection{Learning Braille with Sight: Faster Yet Slower}
Similar to learning as adults, there were tradeoffs of learning braille with full or partial sight. 

%faster
Sighted participants reported many benefits of learning by vision over touch. For many (n=5), this was simply by virtue of having extensive experience learning by sight and comparatively little through tactile sensory channels: 
\begin{quote}
    \textit{“We don’t have that [tactile] sensitivity. So it’s a lot easier for us to continue using our sight to learn. Because, we already have that sense and that’s already how we interact with the world”} (P7).
\end{quote}
Sighted learners reflected that braille tactilely meant \textit{“learning it [in] two different ways”} (P7), which took \textit{"a long time"} (P9) and \textit{"a lot of [finger] scrubbing"} (P10). Factors such as chronic health conditions (e.g., diabetic neuropathy) (P9), aging (P10), and finger calluses (P11) could decrease fingers’ sensitivity, making tactile training even less appealing. Surprisingly, even P11, a low vision braille learner, found visual braille easier to learn, since print could be enlarged and studied through sight: \textit{“With print [braille characters], you can enlarge print. But with [tactile] braille, it’s one size only. And it’s hard when it’s so small, to be able to… feel it with your fingertips.”} (P11) 
%even of our \textbf{sighted} participants felt that not having to train their fingers for sensitivity was a positive affordance of learning visually.

%slower
However, at the same time that learning with sight was faster, it was also a slower and more protracted process due to limited passive exposure and reinforcement opportunities. %Despite intensive training, TVIs often struggled to retain braille knowledge due to the complexity of multiple codes, limited reinforcement opportunities, and insufficient passive exposure, making braille both difficult to master and easy to forget.

Not surprisingly, part of the reason learning braille was slow for our participants was simply that learning braille is difficult.
%; there is a large set of contractions to memorize and abstruse underlying content to learn (i.e., higher-level math). %Previous research has documented that there are around 180 contractions in UEB, and the complex rules of applying them may cause confusion in spelling \cite{Englebretson2023}. %I'll replace it with some better citations tomorrow! 
Our participants experienced difficulty learning the rules and contractions of literary braille (n = 14). Some TVIs also found the mathematics and science-related braille (e.g., Nemeth) hard, due to the lack of academic knowledge and the need to shift between codes (P6, P12). For example, P6 thought having knowledge of higher-level math would help him provide accurate math materials to his braille students. However, his math knowledge \textit{"kind of ended in Algebra 2"}, which hindered his learning of Nemeth and math UEB. Additionally, the expected, challenging shift from print to braille in training courses was compounded by shifts from literary to math UEB (P6, P12): \textit{"It’s a code within a code... in [transcribing print to literary] UEB, you are shifting [from print] to code by the correct braille... but that [Nemeth transcription] is even an extra step, that I think, cognitively, makes it very difficult"} (P6).
%This is from the below paragraph: A prior study has documented the massive number of lessons required for learning contracted braille \cite{Farrow2015}. 

However, TVIs identified the main challenge of learning braille was the timescale on which they learned it. %While TVIs underwent intensive up-front training, many lacked the day-to-day passive braille exposure necessary to  solidify their leaning, resulting in forgetting (n = 6). 
%Beyond reporting prior findings that contractions take significant practice \cite{Farrow2015}, 
%Our participants also reported that learning the technologies used for creating and teaching braille (e.g., Perkins Brailler) was daunting. 
First, when learning braille in formal classroom environments, participants were tasked with acquiring too much content -- not just braille, but also tools such as Perkins Brailler and slate and stylus, in too little time. For example, P9 said \textit{"however long the classes are, it seems like you have to know everything in a short time."} Second, and as a result, once TVIs entered the field, they needed reinforcement to solidify their braille skills. However, compared to younger braille students who generally had \textit{"a lot of worksheets... a lot assignments [to reinforce braille learning]"} (P2) in the course of their everyday schooling, TVIs needed to make time to practice braille: 
\begin{quote}
    \textit{"I am sighted, and I do read print as my main form of learning. So just because I don’t read braille consistently, I do have to dedicate my separate time [to practice braille]"} (P4)
\end{quote}
Working with students was one of the primary ways for TVIs to expose themselves to braille, but they might not always have braille students. For example, P9 thought she was \textit{"very fortunate"} to have braille students, so that she could \textit{"keep up"} braille skills (P9). In contrast, P13 was \textit{"unfortunate"} to have no student over the summer break, resulting in him \textit{“already forgetting [braille]”} (P13). P3 mentioned a colleague who had not taught braille students for 25 years, resulting in complete forgetting. Multiple adult learners emphasized the importance of practicing (P2, P4, P9, P10, P13):
\textit{"If you don’t keep it up, you’ll lose it"} (P10).

\subsection{Strategies and Technologies Used: Never Quite Proficient}

% \begingroup
% \renewcommand{\arraystretch}{1.2}  
%%% Uncomment 2 lines above and the line right below \end{table} to adjust line heights %%%
\begin{table}
    \caption{Participant learning configurations}
    \label{tab:configurations}
    \begin{tabular}{l p{16em} p{8em} c } % value in parenthesis defines column width
        \toprule
        \textbf{ID} & \textbf{Started Learning by}& \textbf{Technologies Used}& \textbf{Braille Learned}\\ \midrule
         P1 & Self-taught& BrailleTranslator, AI (ChatGPT) & UEB, Nemeth\\ \midrule
         P2 & Self-taught, then enrolled in TVI's credential program& UEB Online, UEB Math Tutorial, BrailleBlaster & UEB, Nemeth\\ \midrule
         P3 & Enrolled in TVI's credential program& Braille Brain, International Council on English Braille & UEB, Nemeth\\ \midrule
         P4 & Enrolled in TVI's credential program& Braille Brain & UEB, Nemeth\\ \midrule
         P5 & Enrolled in TVI's credential program& Braille Academy, AI & EBAE, UEB, Nemeth\\ \midrule
         P6 & Enrolled in TVI's credential program& UEB Online, Braille Brain, Nemeth Tutorial, BrailleBlaster, BrailleTranslator, Braille Tutor, AI & UEB, Nemeth\\ \midrule
         P7 & Self-taught, then finished a braille transcriber certification program, finally enrolled in TVI's credential program& Braille Brain & UEB, Nemeth\\ \midrule
         P8 & Enrolled in TVI's credential program& UEB Online, Braille Academy & UEB, Nemeth\\ \midrule
         P9 & Self-taught, then finished a braille transcriber certification program, finally enrolled in TVI's credential program& UEB Online, AI (Gemini) & UEB, Nemeth\\ \midrule
         P10 & Enrolled in TVI's credential program& BrailleBlaster, Duxbury, AI & UEB, Nemeth\\ \midrule
         P11 & Self-taught, then enrolled in TVI's credential program& International Council on English Braille, Braille Academy & UEB, Nemeth\\ \midrule
         P12 & Enrolled in TVI's credential program& ABC Braille, Braille Academy & UEB, Nemeth\\ \midrule
         P13 & Self-taught, then enrolled in TVI's credential program& Braille Brain, Nemeth Tutorial, Duxbury & UEB, Nemeth\\ \midrule
         P14 & Enrolled in TVI's credential program& International Council on English Braille& EBAE, UEB, Nemeth\\
         \bottomrule
    \end{tabular}
\end{table}
% \endgroup

\subsubsection{Patchwork Learning}
TVIs pieced together braille expertise through a patchwork of formal training, digital tools, personal strategies, and limited peer support, relying on persistence and resourcefulness to sustain their learning in often isolated conditions.

TVIs reported a mixture of \textbf{formal} and \textbf{informal} braille learning experiences (Table \ref{tab:configurations}). Most of our participants (all but P1) engaged in formal training and certification programs, including TVI's credential program offered by state universities (n = 13) and Massive Open Online Courses (MOOCs) with certificates (e.g., NFB braille certification program for transcribing \footnote{https://nfb.org/programs-services/braille-certification}) (n = 2). TVIs also used a variety of digital tools to support their braille education, both for self-learning and independent navigation and practice after formal training. Commonly used tools included online tutorials (e.g., UEB Online \footnote{https://uebonline.org/}, Braille Brain \footnote{https://braillebrain.aphtech.org}, Nemeth Tutorial\footnote{https://nemeth.aphtech.org}, and UEB Math Tutorial\footnote{https://uebmath.aphtech.org}) (n = 8), braille translators (e.g., BrailleBlaster\footnote{https://www.brailleblaster.org}, BrailleTranslator\footnote{https://www.brailletranslator.org}, Duxbury\footnote{https://www.duxburysystems.com}, and ABC Braille\footnote{http://abcbraille.com}) (n = 5), online rule-books (e.g., International Council on English Braille \footnote{https://www.iceb.org/}) (n = 3), AIs (e.g., ChatGPT and Gemini) (n = 5), and mobile apps (e.g., Braille Academy\footnote{https://brailleacademy.com} and Braille Tutor\footnote{https://apps.apple.com/us/app/braille-tutor/id878463116}) (n = 4). To solidify memory, adult learners applied a variety of approaches, such as memorizing cheat sheets (n=7), taking notes and making flashcards (n=3), checking braille signs from the surrounding environment (n=2), reading (digital or physical) rule books (n=2), and typing with a braille keyboard for online chatting (P2).

%From paragraph below. It was redundnat with quote being used: And P10 described her thought when doing drill-and-kill: \textit{"it’s just like, 'if I gotta do it, I gotta do it'. Get out of my way [laugh]"}.

TVIs engaged in various strategies to support their learning and practicing. Among those, \textit{\textbf{"drill-and-kill"}} (P5, P10) was an effective learning strategy for braille, yet it required endurance and persistence. In the classroom context, P4 shared: 
\begin{quote}
    \textit{"Like I’m typing out a whole page in braille, and I make a mistake in the middle of the page. Then my teacher would made us restart the whole page over again… But that really helped me get that practice and learn"} (P4).
\end{quote}
Post-certification, those TVIs with braille students would \textbf{practice on-the-job} when working with their students or on transcription tasks (P3, P4, P6, P7, P8, P13, P14). 
\begin{quote}
    \textit{"[when] a student notices I made a mistake, 'Oh, you made a mistake here', and I'm like, 'Yes, I did. Good! Thanks for checking me!'"} (P13)
    
    \textit{"I used my [braille transcribing] job as my practice, which I did braille every day... 1-3 hours a day."} (P7)
\end{quote}
In addition, throughout the day both on- and off-duty, TVIs utilized \textbf{fragmented moments} to practice braille: \textit{"when I had a few minutes, like I was in a waiting room or something, I could pick it up and practice... Just trying to squeeze in and get as much training as possible..."} (P10). Last, some TVIs would reach out to the community for help while encountering unfamiliar braille (n=5). However, the utility of this strategy was constrained by the small networks available to TVIs, who often relied on only a few contacts—most commonly braille teachers from training programs (P6, P10, P14) or colleagues and classmates (P1, P7, P10, P11):
\begin{quote}
    \textit{"Part of being a TVI is, you're kind of alone... You're with regular education teachers in a big wide [school district] area. You have to learn to solve your own problems, and that has always been my thing."}
\end{quote}

\subsubsection{Inadequate Learning Resources}
Limitations and issues within the previously mentioned learning resources created barriers for adult braille learners. Though the \textit{drill-and-kill} approach was effective, even highly-self-motivated TVIs found it \textbf{disengaged} (n = 7); with all the \textit{"repetition"} (P7), braille learning was \textit{"tedious"} (P4) and it was \textit{"hard to make it fun"} (P5). This was true even of interactive digital learning resources; in P7's words, \textit{"It's [digital braille learning material] all like bare-bone"} (P7). Further, having to cobble together resources to cover various braille skills was time-consuming (P1, P10). Both P1 and P10 described their experiences searching for certain braille rules — starting with cheat sheets, then Google, and finally turning to "giant" rule-books.

Digital resources used for self-learning were \textbf{unreliable}-lacking coverage, accuracy, functionality, usability, and even accessibility. Some TVIs reported a near absence of resources for specialized codes, like Nemeth math and chemistry notations, which are essential for higher-level STEM classes (P3, P6). Moreover, the resources that were available across codes were often inaccurate (n = 6). P6 found the online tools for practicing braille occasionally adopted an incorrect or rarely used interpretation, requiring the users to input something different from what they had learned to continue the practice: \textit{"...you’re like, 'that’s not correct'. But in your head, you’re like, 'that’s what the program says is correct'."} The online braille translators were not consistently accurate, either: \textit{"I would say if I put like a teacher percentage in it [online translators], some would be [correct for] like 7 to 8 out of 10 of the time"} (P6). TVIs tended to ask AI braille-related questions for its efficiency. They found AI good at redirecting them to correct resources, but lack accuracy in translating braille (P5, P6, P7, P12, P13):
\begin{quote}
    \textit{"Asking it to transcribe the braille isn’t accurate... And it usually doesn’t know Nemeth-specific questions... If I ask about a specific rule that would be written on a website somewhere... and then the AI has a link to wherever they found the rule from, which is actually more helpful."} (P7)
\end{quote}
Participants also expressed their concerns about AI's accuracy on braille due to issues within training data (P6, P12):
\begin{quote}
    \textit{"I’m still hesitant... it is a very unique code, and a small number of people use it... AI works off us [braille-related users] feeding in information. The 'us' isn’t probably that many."} (P6)
    
    \textit{"There’s already a lot of braille out there [on the internet] that is incorrect, so I can only imagine it [AI] learning incorrect braille."} (P12)
\end{quote}
As a result, while a few of our participants still used AI for braille occasionally, participants expressed a distrust and need for care  (P5, P6, P10, P13). For example, P6 would double-check with other braille users to confirm AI's answer. P5, as both a TVI and a braille teacher of pre-service TVIs, warned her students, \textit{"you’re going to be full of errors [using AI]. And you’re not actually learning braille. You’re learning how to copy braille".}

The resources also commonly lacked functionality, usability, and accessibility. Regarding \textbf{functionality} limitations (n = 5), multiple participants noticed that most tools focus on isolated contractions, with limited sentence-level translation (P7, P11, P14). Additionally, translation tools do not offer feedback on mistakes, implicitly assuming that users already know the rules (P3, P4, P11): \textit{"We don’t know why we’re wrong... It [the system] won’t explain... If you don’t know the code or you don’t know the rules, it’s challenging"} (P3). Regarding \textbf{usability}, P9 pointed out that some apps lacked basic error recovery features, such as backspacing and deleting. Ironically, participants reported \textbf{accessibility} issues within braille learning tools. Multiple participants mentioned the non-adjustable screen display settings (P7, P10, P11): \textit{"I am photosensitive, and I need to be able to change my screen so that it can be darker to help with my eyes. And I wasn’t able to do that with any of the apps"} (P7). P11 also found \textit{"the font size aren’t accessible"} in many tools. Additionally, three participants reported that many tools were not screen reader accessible: \textit{"There are four people in my class that were screen reader users. And one of the programs wasn’t screen reader friendly"}(P7). TVIs wanted to be able to use these tools to teach themselves and their students, so they underscored that accessibility is essential.

\subsection{Imagined Future Designs}
Participants imagined mobile phone apps with features that complemented their learning and practicing styles as sighted adults.

\subsubsection{Utilize Fragmented Time for Practice}
Adult braille learners desired a tool that would motivate them to \textbf{"\textbf{multitask}"} and utilize the fragmented time in between daily activities to practice braille (n = 7):
\begin{quote}
   \textit{ "...most of us have so much going on as adults that...we need something to be able to multitask with. Like we’re cleaning... going to and from work... sitting and watching TV... you can do something on your phone... We need something that’s quick and engaging, that will motivate us to do more, but also something that we can just do while we’re doing something else and be able to get the job done."} (P7)
\end{quote}
Additionally, adult learners wanted to have \textbf{short} practice sessions (e.g., 10 to 15 minutes) to keep them engaged (P7, P8, P9, P10, P12, P13): \textit{"Making it an hour, people probably won’t want to do [practice] it for an hour. But if you make 10 to 15 minute lessons, they’re going to want to do that. And then, if they want to do more, they will do more."} (P12)

They preferred \textbf{mobile devices}, as they are portable and frequently on-hand (P7, P8, P11, P14), whereas practicing on devices such as the Perkins brailler was laborious and tied to a specific location; \textit{"[I have to] get out the brailler, set it up... put the paper in..."} (P8). Some additionally suggested learning apps that leverage lightweight, wireless braille displays, such as a Hable \footnote{Example of a Hable: https://www.iamhable.com/en-am/products/hable-one-keyboard}, for practicing braille typing (P8, P10, P14), though they acknowledged tradeoffs of this approach. P14 found typing skills transferable to visual memory: \textit{"When I learned code in my fingers, it was easier to remember the braille configurations mentally. It was a combination effect for me".} However, P10 pointed out that transitioning from vertical Hable to horizontal Perkins could be confusing: \textit{"But the fact that it’s this way [mimic typing on six dots vertically] might be a little daunting for some older adults who may need to learn it this way [mimic typing on six dots horizontally] on a Perkins [brailler]."}

\subsubsection{Support Collaborative Access, Expert Use, and Corrective Feedback}
 TVIs thought the ideal tool should be accessible, usable, and reliable (n = 10). TVIs wanted the app to be \textbf{fully accessible} (i.e., be accessible by magnification and screen reader users) to support the varying visual abilities of themselves and their students (P5, P7, P9, P10, P11). For example, P11, who had low vision, wanted to have auditory description features: \textit{"because then I was able to visually [see the practice questions] and auditorily hear them saying, 'okay, here are dots 1, 4, 5. What letter is it?' "} (P11). For P7, who was photosensitive, apps should enable users to customize their screen lightness. Additionally, P5 and P9 sought tools that support collaborative access between teachers and braille learning students:
\begin{quote}
    \textit{"So many tools... don’t support screen readers, or they don’t support VoiceOver on an iPad... As a TVI and as a university instructor [of TVIs]... I have to be able to use it with everyone, or I can’t use it with anyone."} (P5)
\end{quote}
%TVIs also wanted the app to have customizable screen display settings (P7). 
Participants also wanted learning aids that have \textbf{adjustable difficulty} level (P12) to accommodate the proficiency levels of various users: \textit{"If there was a way to get tested at the beginning like, 'Oh, you already know the alphabet…You can already just practice Grade 2 [braille]'"} (P12).

As TVIs often learned and practiced independently, they wanted the app to give instant \textbf{corrective feedback} and the rule-based reasoning in response to mistakes, to reduce the burden of checking their various reference materials (P2, P3, P8, P11, P13)—\textit{"It could tell you if you’re right or wrong, right away"} (P8) and \textit{"gives you an example"} (P13).

\subsubsection{Gamification and Community-building}
Participants desired gamification and community-building features, to provide intrinsic and extrinsic learning incentives (n  = 9). Participants believed gamification is effective—not only for engaging children, but also adults (P3, P4, P7, P8, P9, P10, P11, P12, P13): \textit{"It [gamification] does for us adults, too, because we’re all kids at heart... It helps get us excited about doing things"} (P7). They imagined games similar to Duolingo with motivational features (P7, P12). For example, P7 identified herself as \textit{"a chronic Duolingo user"}. She liked the features of \textit{"gamified daily lesson"}, thinking they were \textit{"super cute"}. She was also attracted by the motivators which kept her \textit{"coming back to using it every day"} (P7). Participants also imagined children's typing games as a potential approach to practicing braille contractions and 6-key typing (P8, P10): \textit{"It [a typing game] worked with our kids. We had our kids all learn [QWERTY] typing, which is pretty similar. It’s [braille] just a tactile representation of the alphabet"} (P10).

Given the fact that TVIs were isolated from each other to some degree, P7 wanted an online platform that allowed interactions between other adult braille learners around the same learning stage. This could help to connect TVIs together, enhance communication, reduce the sense of isolation, and keep people engaging: \textit{"For the online [classes]… it [would be] so much easier to do it with a class you’re interacting with people, even if it was just to ask a question here and there, or they’ll make some random comment here and there. Just to kind of break up the tediousness... and help to keep engagement"} (P7).

\section{Discussion}
Sighted adult braille learners encountered challenges finding time to learn and practice braille, Our findings suggest that sighted adult braille learners learn braille on substantially different timescales than blind adolescents. Visually impaired adolescents who learn braille typically do so via regular one-to-one training sessions with braille specialists spanning years of instruction \cite{Roe2014}. While adult learners in our study also had access to formal braille learning through TVIs credential programs, instruction took place in an online classroom setting and typically lasted about one year. The curriculum compressed extensive braille content-including practice with related tools like the Perkins Brailler and slate and stylus—into short timeframes, leaving participants feeling rushed and overloaded during instruction. To maintain their braille skills, sighted adults found themselves faced with integrating self-learning into schedules already tight from work and family responsibilities. Yet, current tools for practicing braille require access to inconvenient or heavy devices, such as laptops or the Perkins Brailler, that are not readily available when on-the-go. As a result, many of our participants desired learning tools that \textbf{take advantage of their fragmented spare time.}

Adult learners in our study also reported limited passive braille exposure because they can read and write printed text with sight and rarely had cause to practice braille in the course of everyday reading tasks, leading to quickly forgetting concepts. Further, TVIs reported that having a braille student enabled them to practice their braille skills on-the-job. Yet, TVIs sometimes went long stretches—from months to years—without having a student who reads braille in their caseload. In contrast, when adolescents learn braille, in addition to daily braille reading and writing in their general education classes, one-on-one lessons with braille  specialists often include questioning, listening, and speaking activities to reinforce their learning \cite{Roe2014, Barclay2010}. As a result, adult learners reported that they had a harder time retaining braille learning gains. Unfortunately, existing learning technologies tend to require the TVI to be seated at a desk or in front of a computer, with their full attention on the learning task. Participants envisioned technologies that could create learning opportunities that \textbf{integrate into their regular daily activities}-while cleaning, commuting, or watching TV.

Sighted adults accelerated their braille learning by skipping tactile skills training, but unengaging tools left the process tedious and undermined their motivation. Visually impaired children engage in pre-braille skills training to increase fingertip sensitivity before advancing to literacy \cite{lee2021}, whereas our participants bypassed this phase by using sight, in order to save time and because of physical limitations. However, TVIs’ learning process is still fraught with unengaging and monotonous tools, such as bare-bones learning apps and rulebooks. In contrast, similar tools designed for teaching braille literacy to children include features to keep their attention, such as games \cite{Gadiraju2020, Milne2014, Ara2016, Nahar2015} and carefully-designed curricula from braille-literate TVIs to make braille learning interesting \cite{Willis2007}. Our participants expressed wanting \textit{“super cute,”} (P7) \textbf{\textit{“gamified,”} \textit{“engaging”}} (P7) learning tools that would help \textit{“motivate}” (P7) and \textit{“excite”} (P7) them. 

TVIs lacked adequate resources and support for their own braille learning, both in tools and in human networks. Visually impaired students continuously receive guidance and help from others, primarily TVIs, paraeducators, and parents \cite{Roe2014, Barclay2010, Harris2011, Winantyo2019}. However, the TVIs-in-training in our study reported lacking adequate support and resources for their own braille learning. The braille practice tools they used only flagged errors without explaining the rules behind them, leaving TVIs confused. Additionally, with few connections to turn to for help, they often felt professionally isolated and forced to navigate learning independently. Participants expressed a need for tools that provide \textbf{meaningful feedback}, not only flagging errors but explaining the correct rules with examples. They also envisioned \textbf{community-building} features to connect with other adult braille learners, which they felt could reduce isolation and expand the sharing of resources for their learning.

Learning technologies which apply micro-learning strategies during fragmented time could potentially solve many of the braille learning challenges identified by TVIs. Prior work on micro-learning, originally devised for secondary language acquisition, shows that distributing brief activities across daily routines provides consistent opportunities for practice without the brief, momentary bursts of motivation needed for dedicated study sessions \cite{Gassler2004}. Wait-learning extends this by detecting natural waiting moments (e.g., waiting for an elevator) and offering low-friction micro-exercises that minimize disruption to ongoing tasks \cite{Carrie2017}. Many of the techniques used in these concepts map directly to the challenges our participants described. In prior work on language acquisition, micro- and wait-learning triggered short, adaptive drills at device re-entry or during natural pauses in daily life, providing varied practice without requiring learners to initiate study \cite{Gassler2004, Carrie2017}. These same features map directly to the challenges reported by adult braille learners, who described fragmented schedules, little built-in feedback, and unengaging tools. Foundational papers on micro-learning also argue for repeated encounters with difficult concepts, aligning with prior work finding that repeated exposure leads to increased learning \cite{Dempster1987}, potentially addressing the way TVIs reported forgetting braille conventions after year-long educational programs or limited daily exposure. Together, these parallels align with the gaps identified in our study and suggest that braille learning could be reframed as a set of distributed, micro- and wait-based practices.

Based on these findings, we can imagine braille learning tools for sighted adult braille learners. One such system might allow learners to practice braille while watching TV or streaming,  where a phone listens and presents a braille word from the subtitles or dialogue, and learners type it using a Hable \cite{Hable} or on-screen braille keyboard (Fig. \ref{fig:BackgroundFigure}). Another is wait-learning flashcards, where during natural pauses such as waiting for Wi-Fi or an elevator, the phone presents short, adaptive drills that repeat difficult items more often and provide immediate feedback \cite{Carrie2017}, and progress could be shared with others in a lightweight community feature similar to step-counting apps. A third is feedback-driven tutor tools, which provide instant correctness checks as learners type braille and adapt the next exercise based on mistakes, and could incorporate para-social bonds with tutor figures to reduce the professional isolation TVIs reported \cite{Shihan2024}. Together, these potential systems illustrate how braille practice could be made better structured to the realities of adult braille learning reported by our participants.

\section{Limitations}
Our work had a few limitations worth noting. First, our sample is relatively small compared to other works in HCI conducted on TVIs. Although we found a large proportion of overlapping themes from the last five interviews, and thus decided we have saturated samples, we admit that such a small number would possibly lead to bias. Second, 11 out of 14 of our participants were taught braille by the same teacher, P5, even though they enrolled in TVIs' credential program from different universities in different states. This means that our findings may not apply to all varieties of TVIs' learning experiences. 

\section{Conclusion}
In this study, we conducted semi-structured interviews with 14 adult braille learners with sight about their practices, challenges, and tech needs for braille learning. We identified three findings: (1) advantages and challenges of sighted adult braille learners, (2) inadequate learning resources they adopted, and (3) their vision towards new technologies to support braille learning. According to those takeaways, we highlighted the uniqueness of sighted adult in braille learning and called for mobile-based, gamified technologies with short learning sections to fit into their learning needs. These points lead us to consider future design space to enhance braille learning for sighted adults, and additionally, to facilitate braille literacy of the visually impaired community.

\bibliography{main}

%This defines the bibliographies style. Search online for a list of available styles.
\bibliographystyle{abbrv}

\end{document}